\documentclass[aps,prl,twocolumn,letter,longbibliography,floatfix]{revtex4-1}

\usepackage{graphicx}% Include figure files
\usepackage{dcolumn}% Align table columns on decimal point
\usepackage{bm}% bold math

\usepackage{float}
\usepackage{amsmath}
\usepackage{color}
\usepackage[english]{babel}
\usepackage[noabbrev]{cleveref}

\newcommand{\be}{\begin{equation}}
\newcommand{\ee}{\end{equation}}

\begin{document}

\title{Direct measurement of force configurational entropy in jamming}
\author{James D Sartor, Eric I Corwin}
\affiliation{Department of Physics and Materials Science Institute, University of Oregon, Eugene, Oregon 97403, USA}
\date{\today}

\begin{abstract}
Thermal fluctuations are not large enough to lead to state changes in granular materials. However, in many cases, such materials do achieve reproducible bulk properties, suggesting that they are controlled by an underlying statistical mechanics analogous to thermodynamics.
While themodynamic descriptions of granular materials have been explored, they have not yet been concretely connected to their underlying statistical mechanics.
We make this connection concrete by providing a first principles derivation of the multiplicity and thus the entropy of the force networks in granular packings. 
We directly measure the multiplicity of force networks using a protocol based on the phase space volume of allowed force configurations.
Analogous to Planck's constant, we find a scale factor, $h_f$, that discretizes this phase space volume into a multiplicity.
To determine this scale factor, we measure angoricity over a wide range of pressures using the method of overlapping histograms and find that in the absence of a fundamental quantum scale it is set solely by the system size and dimensionality. This concretely links thermodynamic approaches of angoricity with the microscopic multiplicity of the force network ensemble.
\end{abstract}

\maketitle

\textit{Introduction.} Thermodynamics connects abstract and difficult to measure details, such as entropy, with more easily measured bulk properties, such as temperature. In granular systems, for which the thermal energy scale is irrelevantly small, similar connections have been proposed for the volume ensemble \cite{edwards_theory_1989,bi_statistical_2015} using compactivity as a temperature analog and also for the force network ensemble \cite{edwards_distribution_2008, bi_statistical_2015} using angoricity. 
While these quantities are measurable \cite{puckett_equilibrating_2013,bililign_protocol_2019}, they are not physically meaningful unless they 1) are shown to have temperature-like properties, such as following the zeroth law and 2) can be rigorously linked to a first principles definition of microscopic entropy~\cite{grimus_100th_2013}. 
Entropy itself was initially an empirical quantity until Sackur and Tetrode placed it on firm footing for the ideal gas with the discretization of phase space into quantum mechanical states \cite{tetrode_chemische_1912,grimus_100th_2013}. The length scale of the discretization depends both on properties of the system and the universal constant $\hbar$, whose value cannot be inferred from bulk properties of the ideal gas alone.
Angoricity holds promise as a temperature analog, as it has been shown to follow the zeroth law, while compactivity fails to do so \cite{bi_statistical_2015, puckett_equilibrating_2013,edwards_granular_2002}. However, before the thermodynamic approach of angoricity can be considered to be on solid ground, the nature of the entropy of jammed systems must first be understood. 

When the density of an overjammed packing increases, force networks are affected in two ways: 1) force magnitudes, and thus pressure, increase, and 2) new contacts between particles form, increasing the number of contact forces in the network. Both of these changes increase the entropy of the force networks. While the effect on entropy from  pressure changes is well understood \cite{bililign_protocol_2019,henkes_statistical_2009}, the effect from changes in the contact network is not. To decouple these effects, we propose an extension to the Force Network Ensemble in which changes in the contact network are allowed. This leads us to identify a critical number of excess contacts, $\delta z_c$, describing the transition from a regime in which entropy is dominated by changes in pressure to one in which it is dominated by changes in the contact network.

The temperature analogue angoricity is defined as the derivative of entropy with respect to the stress tensor
\cite{edwards_distribution_2008}. In isotropic systems this tensor quantity can be simplified to a scalar derivative of entropy with respect to pressure. Just as temperature of an ideal gas can be measured from the velocity distribution, angoricity can be measured from the distribution of local pressures \cite{bililign_protocol_2019}. As a derivative, angoricity provides information about the difference in entropy between two systems but not the absolute values. Previous theoretical and experimental work has identified an inverse scaling of angoricity with pressure in the near jamming limit for two-dimensional (2D) soft spheres \cite{henkes_statistical_2009,bililign_protocol_2019}. However, these studies do not systematically explore the effect of changing the contact network, which remains static in the near jamming limit. In our computational study, we explore the system by varying the spatial dimension, pressure, and number of particles over ranges much larger than would be feasible in a physical experiment.

In this Rapid Communication, we present a first principles derivation of the entropy for the force networks of granular packings.  We measure this entropy up to a multiplicative constant, $h_f$, in the near jamming limit by directly measuring the volume of the space of allowed force configurations.  Analogous to Planck's constant in the Sackur-Tetrode equation, $h_f$ discretizes the space of force configurations into an integer number of accessible states.  We then use the method of overlapping histograms to measure angoricity as a function of pressure, and compare with our force volume measure to solve for $h_f$. This concretely connects the bulk nature of angoricity with the microscopic multiplicity of the force network ensemble.

\begin{figure*}[t]
\centering
\includegraphics[width=\textwidth, trim=150 70 150 70, clip]{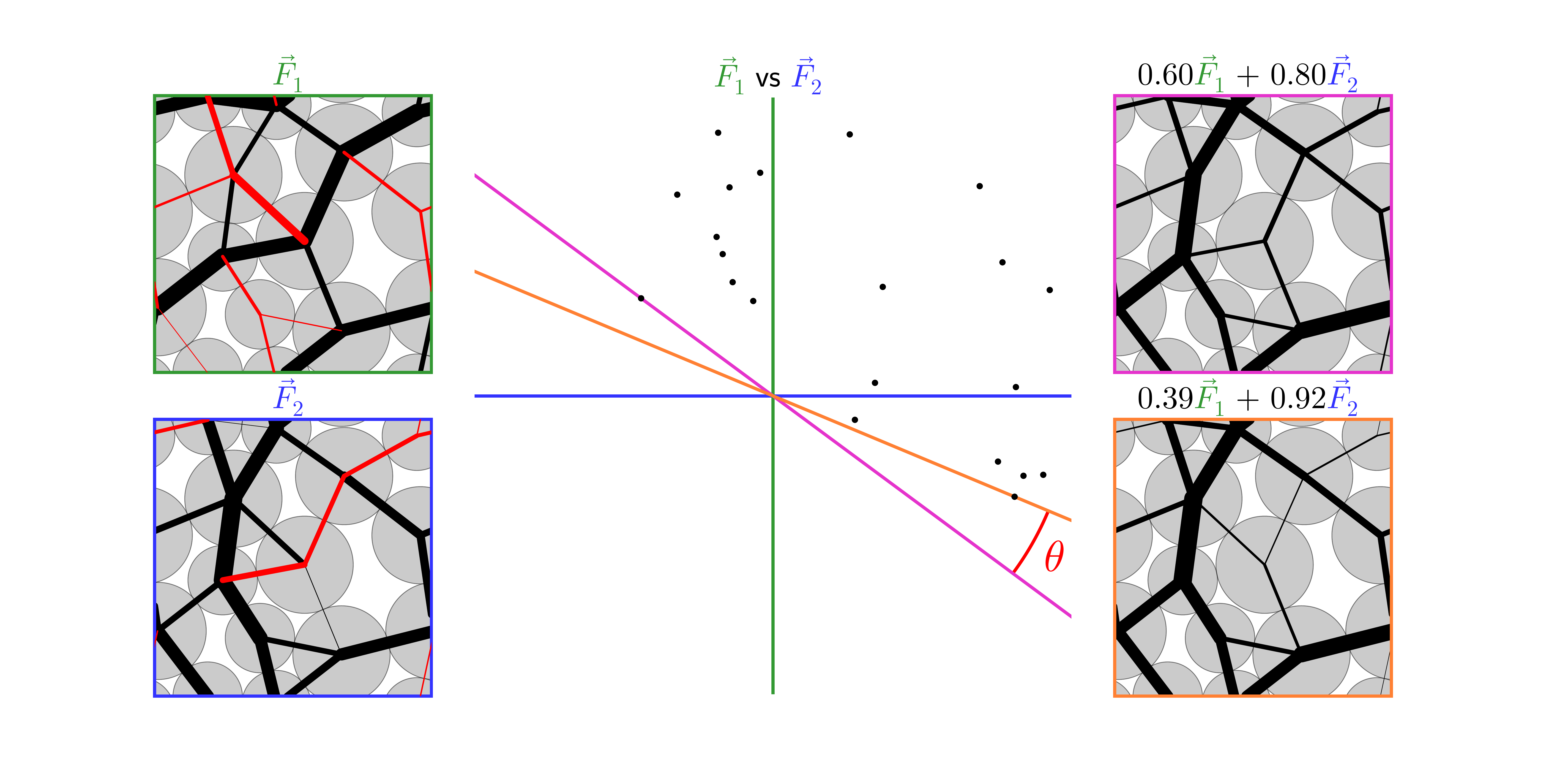}
\caption{Force volume measurement for a system with one excess contact. Left, the two independent states of self stress, $F_1$ and $F_2$. Black lines between particles represent positive (compressive) forces, red lines represent negative (tensile) forces. 
Center, a scatter plot of $F_1$ vs $F_2$ for each pair of particles. Linear combinations of $F_1$ and $F_2$ are represented graphically by drawing a sloped line through the origin and measuring the distance to each point.  Any sloped line for which all of the points fall into the same half-space corresponds to a positive definite linear combination. The set of lines which allow for such solutions is the force space volume, indicated by the angle $\theta$. Note that in a system with $\Delta Z$ excess contacts, this volume is a $\Delta Z$ dimensional quantity.
Right, the two extremal positive-definite linear combinations at the edge of this region are shown. Each has one force brought to precisely zero.}
\label{methodIllustrationFigure}
\end{figure*}

 \textit{Computational methods.} We use pyCudaPacking \cite{charbonneau_universal_2012}, a GPU-based simulation engine, to generate energy minimized soft sphere packings at specified pressures in periodic boundary conditions. We do so for number of particles, $N$, spanning from 256 to 4096, and dimension, $d$, from 2 to 5. The particles are monodispersed, except in 2D in which we use equal numbers of bidispersed particles at a size ratio of 1.4:1 to prevent crystallization. Particles interact through a harmonic contact potential as defined in \cite{charbonneau_universal_2012}, and the system's energy is minimized using the FIRE minimization algorithm \cite{bitzek_structural_2006}.
 
Starting with random initial positions, we minimize energy and then adjust overall density by uniformly scaling particle radii to achieve a pressure $P$ of $10^{-2}$ in natural units, as defined in \cite{ohern_jamming_2003}. This pressure is chosen to prevent crystallization artifacts from high density packings. From there, we iteratively adjust the density both up and down to achieve specific values of pressure.  We do this efficiently by exploiting the known linear scaling of pressure with density above jamming for a harmonic potential \cite{goodrich_scaling_2016}. For each targeted pressure, we ensure that the actual pressure is accurate to a factor of $10^{-5}$. We sample 100 logarithmically spaced steps per decade of pressure to ensure sufficient overlap between the distributions of local pressure for neighboring systems, as is needed for the method of overlapping histograms.

\textit{Rigidity.} To understand the behavior of packings close to the jamming transition we examine the geometric mechanisms necessary for rigidity by constructing an unstressed spring network with the geometry of the packing. The rigidity matrix \cite{charbonneau_jamming_2015, ellenbroek_rigidity_2015, f._hagh_broader_2019}, $\mathcal{R}$, describes this spring network by encoding the normalized contact force vectors from the packing, $n_{i j}$, between pairs of particles $i$ and $j$ as
\begin{equation}
    \mathcal{R}^{k \alpha }_{\left<ij\right>}=(\delta_{jk}-\delta_{ik}) n^\alpha_{ij},
\end{equation}
 where $k$ indexes contacts and $\alpha$ indexes spatial dimensions. For a system with $N_\textrm{stable}$ stable particles and $N_\textrm{contact}$ contacts, this will be an $N_\textrm{contact}$ by $N_\textrm{stable}d$ matrix.  The singular value decomposition of this matrix yields two sets of singular vectors, analogous to eigenvectors for a square matrix. The right singular vectors describe the normal modes of position displacements, and the left singular vectors describe the normal modes of force displacements. 
The left singular vectors corresponding to zero eigenvalues represent mechanically stable force configurations, termed states of self stress.
These vectors need not be positive definite, and therefore are not necessarily valid force configurations for the underlying packing.
 
The magnitude of each contact force can be considered as a degree of freedom while the requirement for mechanical stability introduces $d$ constraints for each particle.  Balancing these constraints requires a minimum number of contacts to ensure stability, which in systems with periodic boundary conditions is given by~\cite{dagois-bohy_soft-sphere_2012, goodrich_finite-size_2012}
 \begin{equation}
     N^\textrm{min}_\textrm{contact} = d(N_\textrm{stable}-1)+1.
 \end{equation} 
A system with this minimum number of contacts has exactly one state of self stress, and each additional contact formed imparts an additional independent state of self stress. Thus, we define the number of excess contacts, $\Delta Z$ as
\begin{equation}
    \Delta Z=N_\textrm{contact}-N^\textrm{min}_\textrm{contact},
\end{equation}
 making the number of independent states of self stress $\Delta Z+1$. We define the number of excess contacts per particle,
 \begin{align}
 \delta z = 2\Delta Z / N,
 \end{align}
where the 2 reflects that each excess contact is shared between two particles. These independent states of self stress form a basis for the $\Delta Z+1$ dimensional space of all mechanically stable force configurations of the spring network. However, imposing a normalization condition restricts this to a $\Delta Z$ dimensional subspace.

\textit{Force volume.} The force network ensemble samples all valid force networks in the spring representation of a packing with equal probability \cite{snoeijer_force_2004,tighe_force_2010,tighe_stress_2011}. 
To determine the force volume, we calculate the normalized independent states of self stress where $F_\mu^q$ is the contact force on contact $q$ in the state of self stress $\mu$.  The set of all possible repulsive contact forces is defined by linear combinations that satisfy
\begin{align}
\label{eqn:coefficientCondition}
\sum_\mu \lambda_\mu F_\mu^q \ge 0
\end{align}
for all contacts $q$, where $\{\lambda_\mu\}$ are coefficients subject to the normalization condition $\sum_\mu \lambda_\mu^2 = 1$.  We define the force volume $V_f$ to be the volume of the space of $\lambda_\mu$ coefficients that satisfy this rule as illustrated in Fig. \ref{methodIllustrationFigure}.  

We measure this force volume with the following protocol:
\begin{enumerate}
\item Recast $F_\mu^q$ into a set, $\{\vec{C}^q\}$, of $N_{\textrm{contacts}}$ vectors containing the value of the force on contact $q$ in each of the $\Delta Z+1$ states of self stress.
\item Planes which pass through the origin and place all of the $\{\vec{C}^q\}$ into a single half-space satisfy inequality (\ref{eqn:coefficientCondition}).  We compute the extremal values of such planes as the facets of the convex hull \cite{barber_quickhull_1996} of $\{\vec{C}^q, \vec{0}\}$.  The normal vector to each facet is the $\{\lambda_\mu \}$ which defines a vertex of the allowed space of coefficients and corresponds to a linear combination of the independent states of self stress in which exactly $\Delta Z$ forces are precisely 0.
\item To respect the normalization requirement we calculate $V_f$ as the $\Delta Z$ dimensional solid angle subtended by the volume defined by these vertices in coefficient space.
\end{enumerate}

We convert this volume into a pure number of configurations by discretizing it into hypercubes of side length $h_f$, named to emphasize the parallelism with Planck's constant $h$ used in the enumeration of phase space states in the Sackur-Tetrode equation.
Because the pressure sets the scale of the average force, we then multiply this enumeration by the pressure, as has been shown in previous theoretical and experimental work~\cite{henkes_statistical_2009,bililign_protocol_2019,mailman_using_2012}. Putting these considerations together, we arrive at an ansatz relating the microscopic force volume to the multiplicity, and thus the entropy:
\begin{align}
\Omega = P \frac{V_f}{(h_f)^{\Delta Z}} && \Longrightarrow && S=\ln P  + \ln{V_f} - \Delta Z \ln{h_f}.
\label{eqn:entropy}
\end{align}

Although pressure and number of excess contacts both appear in the entropy, they are not independent variables but related in the thermodynamic limit by~\cite{ohern_jamming_2003,goodrich_scaling_2016}
\begin{align}
\label{eqn:excess}
 \Delta Z = B(d) N \sqrt{P}.
\end{align}
where $B$ is some function of dimension only. We find values of $B$ of approximately 2.1, 6.0, 12.5, and 23 in dimensions two, three, four, and five. These values are roughly consistent with previous studies for two and three dimensional spheres \cite{ohern_jamming_2003, goodrich_finite-size_2012}. 

Angoricity~\cite{edwards_distribution_2008}, $\alpha$, is derived as:
\begin{align}
\label{eqn:angoricity}
\alpha &\equiv \frac{\partial S}{\partial P} = \frac{1}{P}  + \frac{\partial }{\partial P}\ln{V_f} - \frac{1}{2} \frac{BN}{\sqrt{P}}\ln{h_f}.
\end{align}
First, we measure the volume of force space $V_f$ and explore how it scales with the number of excess contacts. Second, we measure bulk angoricity to confirm our prediction in Eq. (\ref{eqn:angoricity}) and measure the microscopic constant $h_f$.

\begin{figure}[h!]
\centering
\includegraphics[width=\columnwidth, trim=137 240 165 254, clip]{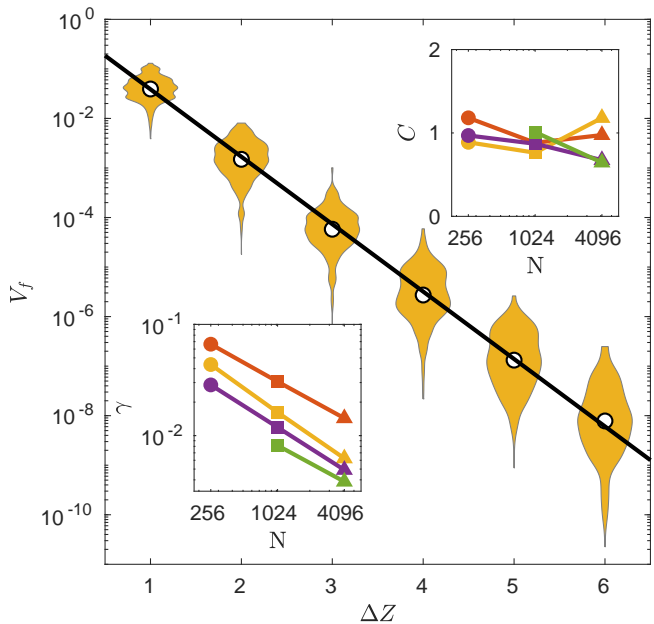}
\caption{Representative exponential scaling of the force volume, $V_f$, with number of excess contacts, $\Delta Z$, for $N=1024$, $d=3$. The median of the distribution for each $\Delta Z$ is shown as a white circle, surrounded by the full distribution in yellow. The black line shows the exponential fitting form, with exponential base $\gamma$. Inset bottom left, $\gamma$ for each $N$ and $d$. Inset top right, the scale, $C$ of the exponential. Inset data is presented for $d$ = 2 (red), 3 (yellow), 4 (purple), and 5 (green), and $N$ = 256 (circles), 1024 (squares), and 4096 (triangles).}
\label{volumePlot}
\end{figure}

\textit{Results.} 
As shown in Fig. \ref{volumePlot}, the measured force volume scales exponentially with the number of excess contacts:
\begin{align}
 \label{eqn:forceVolume}
 V_f~=~C\gamma ^ {\Delta Z}.
\end{align}
We find $C$ to be well approximated by $1$, as shown in the top inset. The lower inset shows that $\gamma$ decreases with increasing $N$ and $d$. 

We can simplify the expression for angoricity by combining the preceding three equations to find
\begin{align}
%\alpha &= \frac{1}{P} + \frac{BN}{2}P^{-\frac{1}{2}}\ln\left(\frac{\gamma}{h_f}\right) \\
\label{eqn:angoricityFittingForm}
\alpha &= \frac{1}{P} + \frac{1}{\sqrt{P_c P}} 
\end{align}
where the crossover pressure between the two power laws is
\begin{align} 
\label{pc_equation}
P_c=\left[\frac{BN}{2} \ln\left(\frac{\gamma}{h_f}\right)\right]^{-2}.
\end{align} 

\begin{figure}[h!]
\centering
\includegraphics[width=\columnwidth, trim=143 240 165 253, clip]{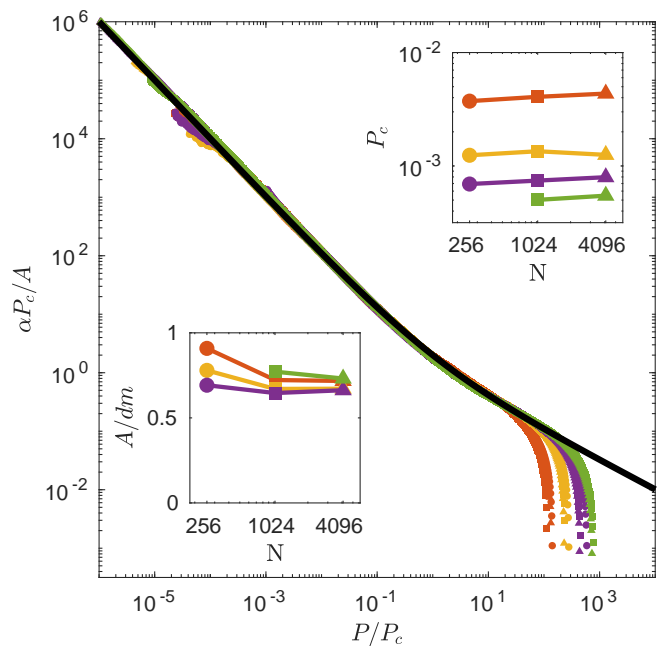}
\caption{Scaled angoricity, $\alpha P_c/A$, for all $N$ and $d$, collapses onto equation \ref{eqn:angoricity} (black line) when plotted against scaled pressure, $P/P_c$, until high pressure deviations caused by second nearest neighbor interactions. Inset top right, the crossover pressure $P_c$. Inset bottom left, $A/dm$ is approximately 0.7. Colors denote dimension from 2-5 and symbol denotes number of particles as in figure \ref{volumePlot}.
}
\label{angoricityPlot}
\end{figure}

We use the method of overlapping histograms of local pressures~\cite{bennett_efficient_1976,mcnamara_measurement_2009,bililign_protocol_2019} to measure angoricity and determine the value of $P_c$ and therefore $h_f$. For each system, we measure the local pressure for many random samples of a particle with its $m=50$ nearest neighbors. The choice of $m$ controls the sharpness of the local pressure distribution and so induces a trivial prefactor $A$, shown in the inset to Fig. \ref{angoricityPlot} to be proportional to $dm$. We then compute the angoricity by comparing these local pressure distributions as in Ref~\cite{bililign_protocol_2019}. We fit the angoricity curve to the power law in Eq. (\ref{eqn:angoricityFittingForm}) with prefactor $A$ and an additive offset. As shown in Fig. \ref{angoricityPlot}, all data collapse onto Eq. (\ref{eqn:angoricityFittingForm}).  We extract the crossover pressures, $P_c$, shown in the upper inset of figure \ref{angoricityPlot}, and find that they are insensitive to $N$, but decrease with increasing $d$.

\begin{figure}[ht!]
\centering
\includegraphics[width=\columnwidth, trim=138 254 176 268, clip]{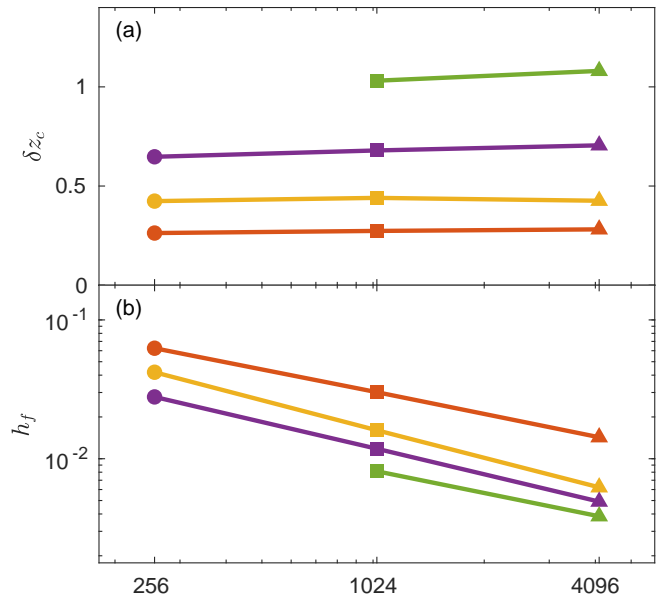}
\caption{Upper, scaling of $\delta z_c$ with $N$ and $d$. Lower, scaling of $h_f$, with $N$ and $d$, calculated from $P_c$ by inverting equation \ref{pc_equation}.  Colors denote dimension from 2-5 and symbol denotes number of particles as in figure \ref{volumePlot}. }
\label{lnhgPlot}
\end{figure}

\textit{Discussion.} --
From Eq. (\ref{pc_equation}) and our measured values of $\gamma$ and $P_c$ we compute $h_f$, shown in the inset to Fig. \ref{lnhgPlot}. A complete expression for entropy can now be written as
\begin{align}
S &= \ln P + \Delta Z \ln\left(\frac{\gamma}{h_f}\right).
\end{align}
This can be recast into a natural form using \cref{eqn:excess,eqn:angoricityFittingForm} by expressing the ratio of $\gamma$ and $h_f$ as a critical number of excess contacts per particle, 
\begin{align}
\delta z_c &= 2B \sqrt{P_c}= \frac{2}{N\ln\left( \frac{\gamma}{h_f}\right)} \\
S &= \ln P + \frac{\delta z}{\delta z_c}. \label{eqn:finalEntropy}
\end{align}
Thus, the entropy is dependent on two intensive thermodynamic variables, $P$ and $\delta z$, and a constant $\delta z_c$ for each dimension. While $h_f$ is observed to decrease with $N$ and expected to vanish in the thermodynamic limit, we find $\delta z_c$ to be intensive with system size, as shown in Fig.~\ref{lnhgPlot}.

The first term in Eq. (\ref{eqn:finalEntropy}) describes the entropy increasing from the absolute pressure scale, whereas the second describes the entropy increasing from the number of contacts increasing. Sufficiently close to jamming the first term will dominate as there will be few changes in the contact network even as the pressure changes dramatically. Further from jamming the second term will dominate, reflecting the primacy of changes in the contact network.
Note that while this equation may be rewritten as a function of pressure using Eq. (\ref{eqn:excess}), for any particular finite packing the integer number of excess contacts is required to calculate the entropy precisely.

\textit{Conclusion.} We have demonstrated that the force network ensemble framework can be used to directly compute the multiplicity of the force configurations in packings close to the critical jamming point. We have presented an ansatz linking the volume of the force configurational space associated with a packing to the entropy of the packing.  This entropy can be expressed as a function of pressure and is independently confirmed by measurements of the angoricity over approximately seven orders of magnitude of pressure.  We have combined these two approaches of measuring entropy in order to extract the fundamental scales governing the discretization of phase space that allows for enumeration. We discover a crossover value for the excess contacts per particle, $\delta z_c$, below which the entropy is governed primarily by changes in pressure at fixed contact network and above which the entropy is governed primarily by the creation of new contact forces.

This work places angoricity on a firm footing as a thermodynamic quantity that controls the behavior of overjammed systems.  By tracing this entropy all the way down to an enumeration of states we discover that, perhaps unsurprisingly, Planck's constant does not set the fundamental scale of discretization $h_f$. In a purely classical model such as this, the discretization can only depend on the finite size effects of the system which are determined by $N$ and $d$. Thus, in the thermodynamic limit, while $h_f$ vanishes, the behavoir of the system is controlled by $\delta z_c$ and thus $P_c$ which do obtain fixed values. This full expression for entropy provides the first concrete linking of the microscopic force network ensemble to the thermodynamic description of granular materials and offers a complete description for the thermodynamics of the force networks in overjammed systems.

\textit{Acknowledgments.} We thank Bulbul Chakraborty, Karen Daniels, Sean Ridout, and Brian Tighe for useful discussions. This work benefited from access to the University of Oregon high performance computer, Talapas. This work was supported by National Science Foundation (NSF) Career Award DMR-1255370 and the Simons Foundation Grant No. 454939.

\bibliography{Angoricity}% Produces the bibliography via BibTeX.

\end{document}